# Thermal transport in van der Waals graphene/h-BN structure: A molecular dynamics study

M. Sadegh Alborzi and Ali Rajabpour

Advanced Simulation and Computing Laboratory, Mechanical Engineering Department,

Imam Khomeini International University, Qazvin, Iran.

Among the van der Waals heterostructures, graphene/h-BN heterostructure is an appropriate candidate for 2D nanoelectronic devices. In this paper, using non-equilibrium molecular dynamics simulation approach, heat transport in bilayer graphene/h-BN and graphene/h-BN van der Waals heterostructure (i.e. h-BN flakes periodically inserted on the top and bottom of a graphene layer) are explored. The results show that by increasing the length of the system, the thermal conductivity of bilayer graphene/h-BN increases. Furthermore, it was revealed that heat transport in graphene/h-BN heterostructure enhances compared to that in monolayer graphene or monolayer h-BN. The size effect analysis shows that the heat fluxes passing through each layer in bilayer graphene/h-BN converges when the size of the system is larger than 100 nm. The results can improve the understanding heat transfer phenomena in the van der Waals heterostructures and improve designing of heterostructures for better thermal management and heat dissipation.

**Introduction**

Thermal regulation has become extremely important in microprocessors over the last two decades as the scale of electronic devices has shrunk. Therefore, thermal management and heat dissipation become vital to the device's efficiency and reliability. To manage heat transport, van der Waals heterostructures (layered combinations of different 2D materials) have prepared the appealing opportunity to reach this aim. At the nanoscale, heterostructures have been shown to perform substantially better than their constituents in terms of reinforcing thermal, electrical, and mechanical properties [1]. Among the van der Waals heterostructures, graphene/h-BN heterostructure is an appropriate candidate to be employed in 2D devices due to the structural similarity of these two materials (1.8% lattice mismatch). Moreover, h-BN, which is known as the only 2D insulator[2], [3], can be utilized to improve the electronic features of graphene-based 2D devices [4].

There are some approaches to control the heat transport and tuning the thermal conductivity of the van der Waals heterostructures, such as interlayer bonding, junctions, and functionalization of the layers. Interlayer sp3 bonding was used to investigate the thermal conductivity of bilayer graphene. It was discovered that 5% interlayer sp3 bonds, which are uniformly dispersed in the bilayer graphene, would reduce the thermal conductivity by up to 70% compared to pristine bilayer graphene. [5]. With the same approach, Iwata and Shintani have investigated thermal conductivity in presence of sp3 bonding in bilayer graphene/ h-BN heterostructure. [6]. They found that increasing the interlayer bonding up to 25%, significantly reduced the thermal conductivity of the bilayer graphene/ h-BN. However, when the percentage of the interlayer bonding exceeds 25%, the thermal conductivity of this structure gradually increases. This behavior of the thermal conductivity could be explained by the local phonon density of states. The interlayer bonds disperse the phonon at low fractions of sp3 bonds, while above the critical percentage (25%) of interlayer SP3 bonds, their contribution to the integration of the two-layer bond is greater, and hence the rigidity of the bilayer structure steadily rises. Furthermore, simulation of the graphene/ h-BN bilayer with $sp^3$ bonding in the presence of nanoholes in the bilayer has proved that the defective graphene layer is mechanically stronger than the h-BN layer [7].

Junctions such as step-like, overlaying, and overlapping are other methods to control the heat transport in heterostructures. In synthesized graphene, the top layer is abruptly disrupted, while the bottom layer continues to function normally [8], [9], which is called a step-like junction. MD simulation of bilayer–monolayer graphene structure has shown a temperature difference between the sheets at the step junction while this temperature jump would eventually vanish as the device length increases. [10]. Moreover, analytical modeling of overlapping [11] and overlaying [12] junctions in a van der Waals heterostructure have shown that growing the span of the overlapping territory, enhances the heat flux passing through the heterostructure. Moreover, simulation of the interfacial heat conductance for an overlapping junction in graphene and h-BN flakes has shown that the conductance values are independent of the contact area [13]. In a recent study, the NEMD approach was used to assess the influence of the coverage percentage and layer configuration on thermophysical properties in bilayer graphene/h-BN. [14]. The authors have shown that increasing the h-BN coverage can improve the thermal conductivity of the heterostructure up to 25%. They also manifested that the ratio of h-BN to graphene sheet has a greater impact on heat transfer than the h-BN layer's geometrical arrangement.

Functionalization of the layer by out of plane bonding with an atom such as silicon [15], hydrogen [16], and nickel [17] can operate as a regulating agent for heat transfer. In graphene/h-BN van der Waals heterostructure, increasing the van der Waals interaction between layers reduces the thermal resistance [18]. It was also found that functionalization of graphene with hydrogen could reduce the cross-plane thermal resistance so that the interfacial resistance decreases to 73% of pristine graphene/h-BN.

Interfacial thermal transport in heterostructures is also an essential factor in heat manipulation. Room temperature (RT) thermal boundary conductance (TBC) for multilayer graphene/ h-BN was reported about 187 MW/m$^2$K [19]. MD simulation of graphene/h-BN nanoribbon has shown that the thermal boundary conductance between graphene and h-BN is around 5 MW/m$^2$K at RT [18]. The stacking arrangement for monolayer graphene sandwiched between h-BN layers on the thermal boundary conductance was also recently investigated [20]. Moreover, using the Raman technique TBC at graphene/h-BN was experimentally reported about 52.2 MW/m$^2$K [21]. TBC for monolayer graphene/h-BN which is located between Ti and SiO$_2$ layers was reported around 18.3 MW/m$^2$K utilizing the time-domain thermoreflectance technique [22].

The bilayer van der Waals heterostructure has been extremely discussed in terms of electronic properties [23], phonon characteristics [24], and thermal properties views [25]. Meanwhile, the aspect of phonon and thermal properties of the heterostructure has attracted considerable attention for nanoelectronics applications [26]. In this regard, bilayer graphene/h-BN heterostructure has the highest value of the thermal boundary conductance in comparison to other bilayer heterostructures. For instance, the thermal conductance of $MoS_2$/h-BN was calculated to be 17 $MW/m^2K$, while this value for graphene/h-BN was measured to be 52.2 $MW/m^2K$ [21]. According to the literature review, the graphene/h-BN heterostructure is a more appropriate choice to design the devices for providing the feasibility of manipulating and engineering heat that it will respond to the growing demands for optimizing heat elimination issues in small-scale devices.

In the present study, thermal transport in bilayer graphene/h-BN is explored at different lengths using NEMD simulations employing optimized Tersoff potential function. To investigate the difference of energy flowing through the graphene and h-BN layers, a two-dimensional energy profile is implemented. Moreover, the thermal conductivity of the bilayer graphene/h-BN is evaluated to probe the length dependence of thermal conductivity. Then, the thermal transport in graphene/h-BN heterostructure is compared with thermal transport in the monolayer graphene, monolayer h-BN, and bilayer graphene/h-BN. Two-dimensional energy profile shows the difference of heat flux passing among these structures. Finally, the effective thermal conductivity of heterostructure is compared with the effective thermal conductivity of monolayer graphene, monolayer h-BN, and bilayer graphene/h-BN.

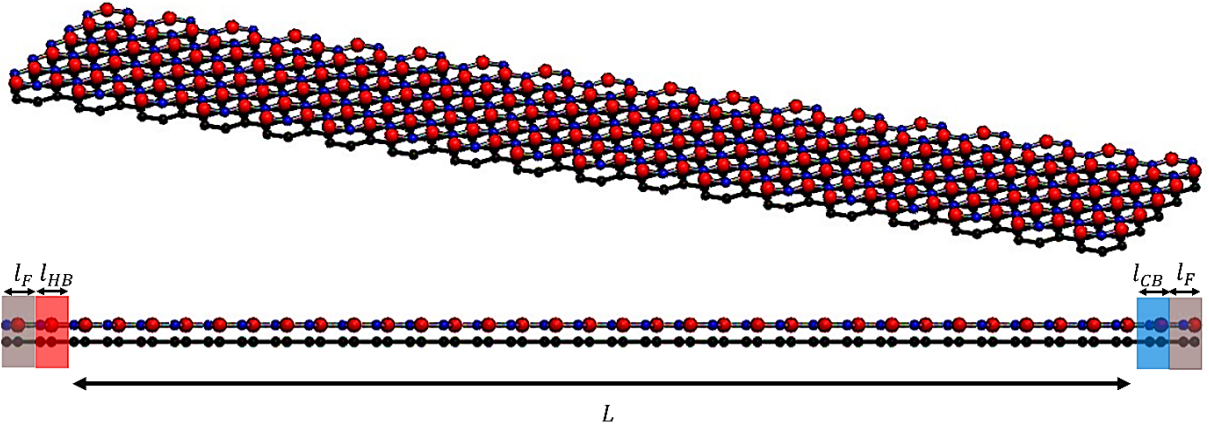

Fig. 1. The atomic structure of the bilayer graphene/h-BN. L introduces the total length of the structure. $l_F$, $l_{HB}$, and $l_{CB}$ denote the length of fixed region, hot and cold baths, respectively.

**Molecular dynamics details**

We utilized non-equilibrium molecular dynamics simulation (NEMD) to calculate the thermal properties of bilayer graphene/h-BN and graphene/h-BN heterostructure. In this approach, insulated boundary conditions were introduced along the x-axis by freezing atoms at two ends of the structure. In the y-direction, periodic boundary conditions were imposed with a thickness of 5 nm, which is wide enough to prevent finite-size effects in the y-direction. The in-plane and cross-plane interactions were defined using the Tersoff potential [27] re-parametrized for graphene by Lindsay and Broido [28] and the Lennard-Jones potential with parameters mentioned in Table 1. Time step of all simulations has been set as 0.5 fs. At the beginning of the simulation, the NPT (constant number of atoms, stress, and temperature) ensemble was applied 0.25 ns by imposing x- and y-direction stresses to be zero. Afterward, NVT ensemble has been employed to relax the structure (excluding frozen atoms) in 300 K for 0.5 ns; then, hot (320 K) and cold (280 K) baths are turned on about 1 ns, and the rest of the structure is imposed to NVE ensemble. Finally, data collection for measuring the temperature distribution in the system began. The framework was divided into multiple statistical slices for this purpose, and Eq.1 was used to measure the temperature (T) in each slice.

$$T = \frac{2}{3K_B N} \sum_{i=1}^{N} \frac{1}{2} m_i v_i^2 \qquad (1)$$

Table 1. LJ parameters for cross-plane non-bonding interactions [6], [29].

| Interaction type | $\varepsilon$ (eV) | $\sigma$ (Å) |
|---|---|---|
| Carbon-Boron | 0.003293 | 3.411 |
| Carbon-Nitrogen | 0.004068 | 3.367 |

Time-averaging over 1 ns was used to measure the final temperature of each slice. Using the Fourier's law and the temperature profile, the effective thermal conductivity, $k$, of the graphene/h-BN heterostructure and bilayer graphene/h-BN can be calculated as follows: (Eq.2):

$$k = q'' \frac{L}{\Delta T} \qquad (2)$$

where $\Delta T$ is the temperature difference between two ends of the system after excluding the temperature jump artefacts. The LAMMPS (Large-scale Atomic/Molecular Massively Parallel Simulator) code was performed to run all of the MD simulations [30].

## 3. Results and discussion

The thermal conductivity of monolayer graphene with a 30 nm length was measured to verify our numerical code. It was obtained about 366 W/mK which is in agreement with earlier stated value of 354 W/mK for the graphene layer [31]–[33]. Table 2, moreover, shows other studies performed to calculate the thermal conductivity of monolayer h-BN and bilayer graphene/h-BN calculated by molecular dynamics and Boltzmann transport equation(BTE).

Table 2. Comparison of the thermal conductivity [W/mK] calculated in this study and previous works.

| Structure | This study | Previous studies |
|---|---|---|
| Monolayer graphene (L=300 nm) | 500 | ~ 530 (MD) [13] |
| Monolayer h-BN | 570 ± 40 | ~ 600 [L= 1 μm] (BTE) [34]<br>~ 600 [L= 1 μm] (MD) [35] |
| Bilayer h-BN/graphene | 1111 ± 50 | ~ 1250 [L= 400 nm] (MD) [35]<br>~ 1000 [L=500 nm] (MD) [6] |

*3.1 heat transport in bilayer graphene/h-BN*

In this section, the bilayer graphene/h-BN structure, as shown in Fig. 1, was considered in order to determine its thermal conductivity. Fig. 2 shows the temperature distribution in bilayer graphene/h-BN at four lengths of 30 nm, 70 nm, 100 nm, and 200 nm. The temperature profiles illustrate that there is no local temperature difference between the graphene and the h-BN layers. To provide a complete perspective of heat transport in this structure, the heat flux passing through the bilayer structure is also illustrated for four lengths of L=30 nm (Fig. 2b), L=70 nm (Fig.2d), L=100 nm (Fig.2f), and L=200 nm (Fig. 2h). The heat flux passing through the graphene and h-BN is shown for each layer separately.

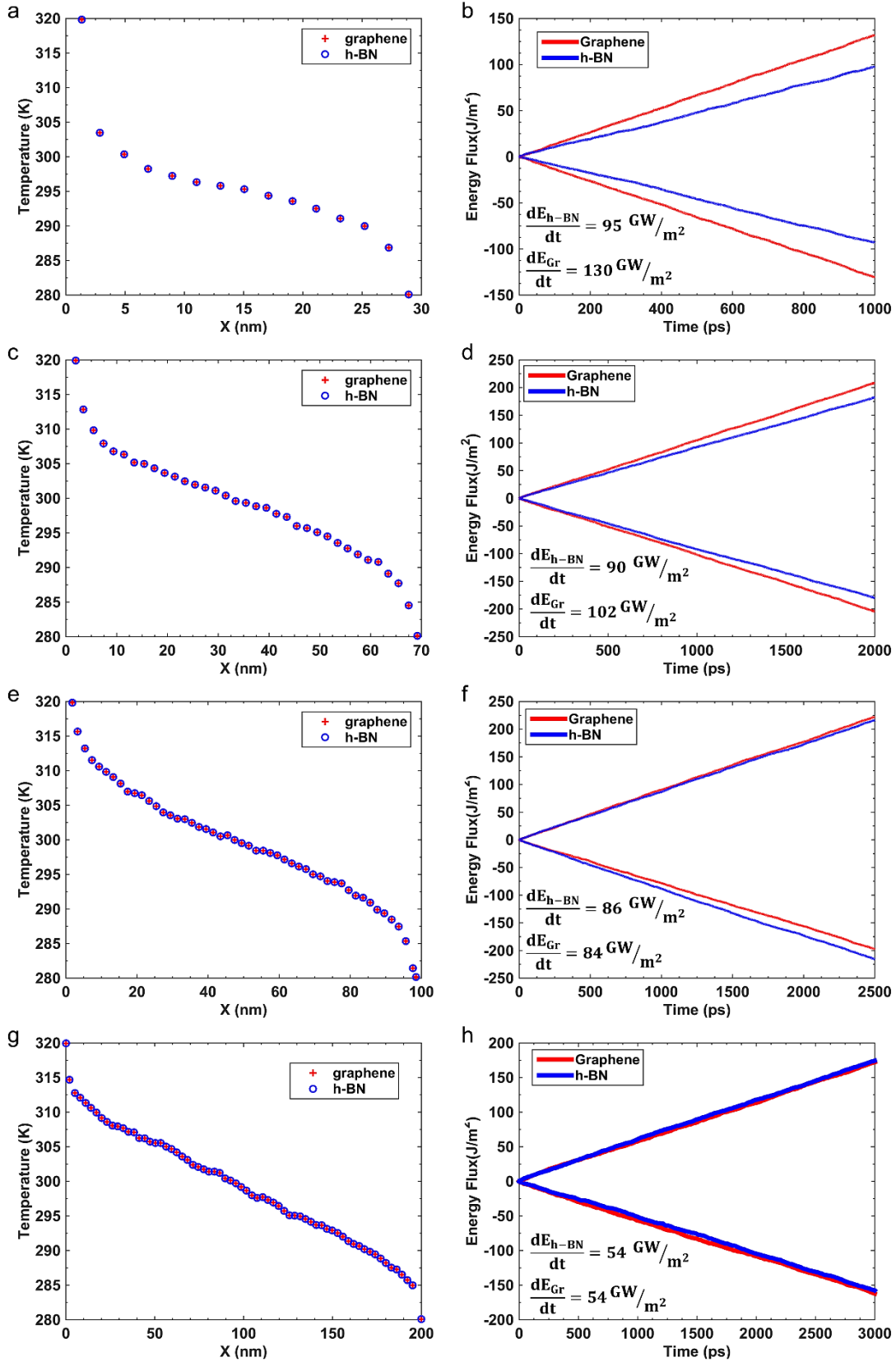

Fig. 2. Temperature distribution and energy flux in bilayer graphene/h-BN with length of 30 nm (a and b), 70 nm (c and d), 100 nm (e and f), and 200 nm (b and d).

It is observed that there is a difference between heat flux passing through the h-BN layer and the graphene layer for L=30nm. Since each layer was connected to a separate heat bath, the graphene layer transmitted more amount of heat flux in comparison to the h-BN layer due to its higher thermal conductivity. The difference between the graphene and h-BN heat flux vanishes when the length of the bilayer structure increases. As seen in Fig. 2h for L=200 nm, the heat fluxes passing through each layer are completely equals. It is observed that the difference between the transmitted heat flux becomes negligible for bilayer structures having a length of more than 100 nm.

The effective thermal conductivity of the bilayer graphene/h-BN can be calculated based on the temperature profile and the heat flux calculated in the previous section. The length-dependency of the thermal conductivity of bilayer graphene/h-BN is shown in Fig. 3a. The results show that the thermal conductivity increases when the length of the system increases. Similar behavior has been shown in graphene/hexagonal boron-nitride polycrystalline heterostructures. The thermal conductivity and heat transfer are tunable by controlling the grain size and percentage of h-BN atoms[36][13]. Recent studies have shown that increasing temperature can reduce the grain size effect on the thermal conductivity of heterostructures[37]–[39]. The thermal conductivity of bilayer graphene/h-BN at infinite length is calculated as 1111 ± 50 W/mK. To make this upward trend tangible, we compared our results with the thermal conductivity of monolayer graphene and h-BN in Fig. 3b. It shows that the thermal conductivity of bilayer graphene/h-BN is less length-dependent compared to that of monolayer graphene. Moreover, it is observed that the thermal conductivity of bilayer graphene/h-BN is more than that of monolayer graphene when the length of the structure is less than 100 nm, and the thermal conductivity of monolayer graphene is more than that of graphene/h-BN bilayer when the length of the structure is more than 100 nm. Furthermore, it is observed that the thermal conductivity of bilayer graphene/h-BN is less than the thermal conductivity of monolayer graphene, and it is more than the thermal conductivity of monolayer h-BN, which means that the thermal properties of this bilayer provide the opportunity to have the system behaving between the monolayer graphene and monolayer h-BN.

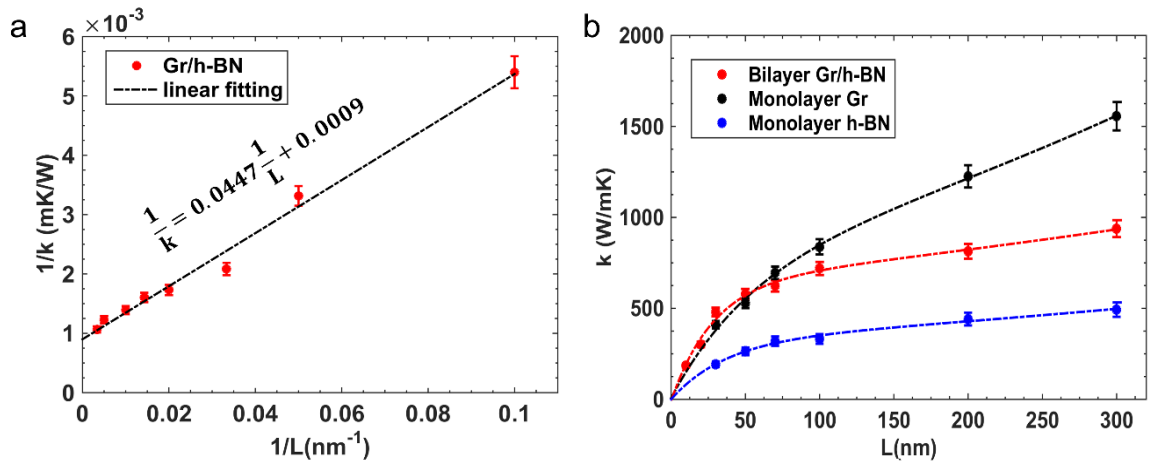

Fig. 3. The inverse of the bilayer graphene/h-BN thermal conductivity versus the inverse of length (a). Comparison of thermal conductivity of bilayer graphene/h-BN with monolayer graphene (b).

In order to investigate whether the thermal conductivity of bilayer graphene/h-BN can be related to the thermal conductivity of each layer, we consider the equivalent resistance circuit for the structure of bilayer graphene/h-BN. Due to the almost zero-temperature difference between the two layers, heat transfer between the two layers can be neglected compared to the heat passing through each layer. Therefore, the equivalent resistance circuit of the structure can be considered as follows:

$$R_{Gr} = \frac{L}{k_{Gr}A}, R_{h-BN} = \frac{L}{k_{h-BN}A}$$
$$\frac{1}{R} = \frac{1}{R_{Gr}} + \frac{1}{R_{h-BN}} \quad (3)$$
$$k = \frac{k_{Gr} + k_{h-BN}}{2}$$

Figure 4 shows a comparison between the thermal conductivity results of bilayer graphene/h-BN and what is obtained from Equation 3. As can be seen, the general behavior of thermal conductivity directly obtained from molecular dynamics and the equivalent thermal circuit are similar. However, the values obtained from the two methods are different. This is because in calculating the thermal conductivity based on the equivalent resistance of the two layers, the heat flux between the layers has been neglected. This is made clearer by reducing the Lennard-Jones potential parameter (*ε*) by half, which reduces the difference between the thermal conductivity of bilayer graphene/h-BN, calculated directly by the molecular dynamics method and by approximating from Equation 3. Obviously, by zeroing the interlayer interaction between graphene and h-BN, the thermal conductivity of both methods will be fully compatible.

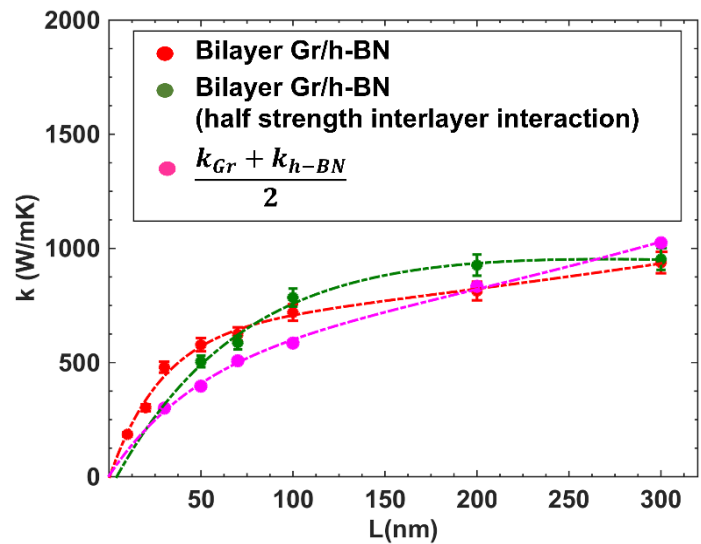

Fig. 4. Comparison of the thermal conductivity of bilayer graphene/h-BN calculated from MD and with half strength of van der Waals interaction between graphene and h-BN and that of obtained from eqn. (3)

*3.2 heat transport in graphene/h-BN heterosructure*

The atomistic structure of graphene/h-BN heterostructure has shown in Fig. 5. As illustrated in this figure, a graphene layer was sandwiched by a specific algorithm of finite h-BN sheets in which the length of a single h-BN was set 6.86 nm. Also, h-BN sheets were placed 0.34 nm above and under the graphene, respectively. In this heterostructure, the graphene layer just connected to the heat bath, and van der Waals interaction is the only force maintaining h-BN flakes.

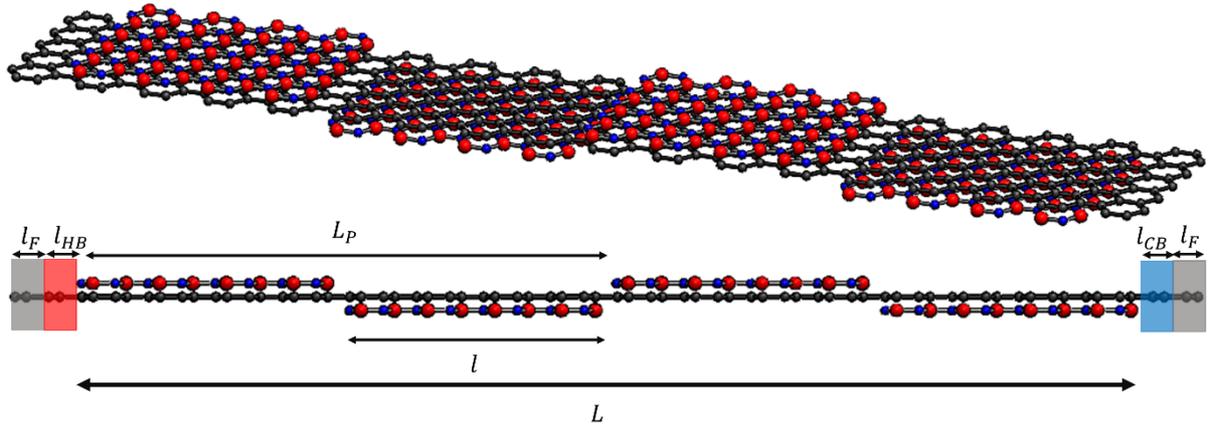

Fig. 5. The atomic structure of the graphene/h-BN heterostructure. $L$ introduces the total length of the structure, $l$ and $l_P$ show h-BN layer length, and periodic length, respectively. $l_F$, $l_{HB}$, and $l_{CB}$ represent the fixed layer, hot, and cold baths regions, respectively.

Fig. 6 shows the temperature distribution of graphene/h-BN heterostructure with a length of 30 nm. Since the h-BN layers were not connected to any heat bath, the temperature of the h-BN layers, which are close to the heat baths is almost constant. Moreover, it is observed that the temperature of the h-BN layers is affected gradually by the graphene layer and could pass an amount of in-plane heat flux.

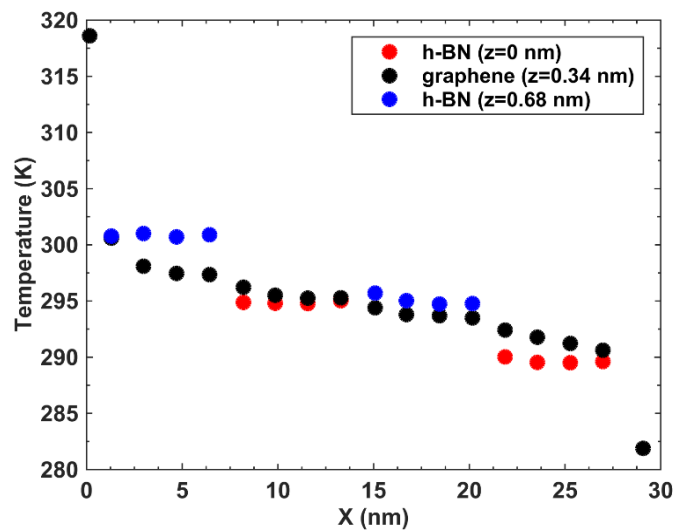

Fig. 6. Temperature distribution in a graphene/h-BN heterostructure with a length of 30 nm.

Fig. 7 depicts the energy flux passing through the heterostructure. In order to illuminate the effect of the h-BN flakes on thermal transport, the heat flux passing through the heterostructure was compared to energy flux passing through monolayer graphene, monolayer h-BN, and bilayer graphene/h-BN. As seen in this figure, the h-BN flakes increases the energy flux passing through the structure in comparison to monolayer graphene, while its value is still less than the heat flux passing through the bilayer graphene/h-BN. This is because in bilayer structure there are two complete layers which the heat flux could passes easily through them, but in the heterostructure there are thermal resistances due to the van der Waals interactions, which reduces the heat flux.

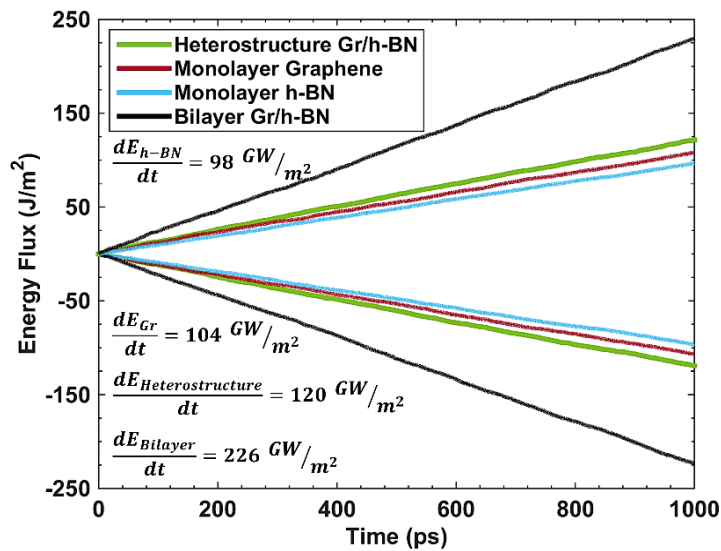

Fig. 7. Comparison of the total energy flux passing through the graphene /h-BN heterostructure, monolayer graphene, monolayer h-BN, and bilayer graphene/h-BN. The length of the all structures are 30 nm.

Furthermore, to provide a complete picture of how the h-BN flakes affect the heat transfer of the heterostructure, the effective thermal conductivity of the heterostructure (Eq.2) was compared to the thermal conductivity of monolayer graphene and monolayer h-BN as depicted in Fig. 8. It is observed that the h-BN flakes have enhanced the effective thermal conductivity of the heterostructure in comparison to the thermal conductivity of monolayer graphene and monolayer h-BN. The effective thermal conductivity of the heterostructure was found about 361 W/mK, which is more than the value of monolayer graphene (321 W/mK) and monolayer h-BN (159 W/mK). Moreover, the thermal conductivity of heterostructure is less than the thermal conductivity of bilayer graphene/h-BN (422 W/mK).

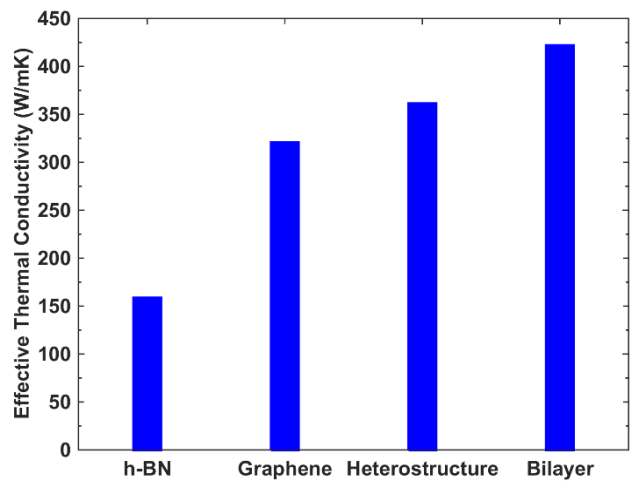

Fig. 8. Effective thermal conductivity of the monolayer graphene, monolayer h-BN, graphene/h-BN heterostructure, and bilayer graphene/h-BN. The length of the all structures are 30 nm.

**Conclusion**

First non-equilibrium molecular dynamics modeling was conducted to study the effective thermal conductivity of bilayer graphene/h-BN. The evaluation of length-dependent thermal conductivity of bilayer graphene/h-BN was made the following conclusions:

- There was no local temperature difference between the graphene and h-BN layer.
- In the case of small lengths, the graphene layer transmitted more amount of heat flux in comparison to the h-BN layer while the difference between the graphene and h-BN heat flux disappeared when the length of the bilayer structure increases.
- The thermal conductivity of bilayer graphene/h-BN was less length-dependent compared to that of monolayer graphene.
- The thermal conductivity of bilayer graphene/h-BN at infinite length is calculated as $1111 \pm 50$ W/mK.

Then, we constructed graphene/h-BN heterostructure using the none equilibrium molecular dynamics simulation in order to compare the heat transport properties of this heterostructure to the graphene monolayer, h-BN monolayer, and bilayer graphene/h-BN, which the following results were extracted:

- The h-BN flakes increase the energy flux passing through the structure in comparison to monolayer graphene and monolayer h-BN, although its value is still less than the heat flux passing through the bilayer graphene/h-BN.

- The effective thermal conductivity of the heterostructure was found about 361 W/mK, which was more than the value of monolayer graphene (321 W/mK) and monolayer h-BN (159 W/mK) and was less than the thermal conductivity of bilayer graphene/h-BN (422 W/mK).

The proposed heterostructure in this study could be considered as an efficient structure to design the devices for providing the possibility of controlling and managing heat.